\documentclass[preprint,10pt]{elsarticle}

\usepackage{amssymb}
\usepackage{amsmath}
\usepackage{placeins}

\biboptions{comma,round,sort&compress}

\usepackage[frozencache,cachedir=.]{minted}

\newcommand{\rzplane}{$Rz$\nobreakdash-plane}
\graphicspath{{images}}

\newcommand{\code}[1]{{\texttt #1}}

\newcounter{bla}

\journal{Computer Physics Communications}

\usepackage{hyperref}

\begin{document}

\begin{frontmatter}

\title{Numerical methods for stellarator simulations in BOUT++}

\author[a]{David Bold\corref{author}}
\author[a]{Brendan Shanahan}

\cortext[author] {Corresponding author.\\\textit{E-mail address:} david.bold@ipp.mpg.de}
\address[a]{Max Planck Institute for Plasma Physics, Teilinstitut Greifswald, 17491 Greifswald, Germany}

\begin{abstract}
  Modeling the Scrape-off layer (SOL) of stellarator fusion devices is
  challenging due to the complicated magnetic topology, requiring
  numerical tools to solve transport equations for realistic
  geometries.
  Previously the flux coordinate independent (FCI) method has been
  successfully applied to model the SOL in simplified geometries.
  The current work presents some of the recent improvements for the
  BOUT++ modeling implemented to simulate the SOL in realistic
  geometries with the example of Wendelstein 7-X.
  The changes include improvements for the grid generation tool, the
  physics model as well as the BOUT++ library itself and are all
  released under an open source license. Several of the features have
  been tested using the method of manufactured solutions. A short outlook
  is given on current modeling work using the new features.
\end{abstract}

\end{frontmatter}

\section{Introduction}\label{s:intro}
Magnetically confined fusion confines a plasma at high temperature in a toroidal magnetic field. The most promising candidates for the realization of fusion energy are axisymmetric tokamaks~\cite{wesson04a}, which rely on an internal current, and stellarators~\cite{helander14a}.
Stellarators are a promising path towards fusion reactors, as they
allow for precise tailoring of the magnetic field to optimize for a
given set of desired quantities. It is beneficial to have both high-fidelity and efficient models to design a stellarator for fusion applications.
Wendelstein 7-X (W7-X) is the most advanced stellarator, and features an island
divertor where the scrape-off layer (SOL) is characterized by several islands that are
intercepted by target plates~\cite{klinger17a,wolf17a,pedersen17a,beidler21a,pedersen22a,pedersen18a,pedersen19a,hammond19a} at discrete toroidal locations.

Understanding and predicting the transport in the scrape-off layer
(SOL) requires accurate numerical modeling of the transport processes
involved. In tokamaks, there exist many high-fidelity codes and models to accurately predict the heat and particle flux in the SOL~\cite{dudson09a, dudson24a, stegmeir15a, ricci12a, tamain14a, wiesen15a}. The complicated edge topology in W7-X, however, inhibits the numerical implementation of these higher-fidelity models which typically rely on a field-aligned approach to modeling, where the evaluation of parallel gradients is facilitated by aligning a numerical coordinate to the magnetic field direction. In regions of magnetic islands and chaotic field lines such as what is found in the W7-X SOL, this method is untenable. As such, transport codes such as EMC3-Eirene~\cite{feng14a,feng21a} are used to model the SOL of stellarators~\cite{winters21a}.  While transport codes are relatively computationally cheap, they rely on the input of diffusion coefficients $D$ and $\chi$. A higher-fidelity model will help to inform the prescription of these diffusion coefficients and elucidate the nonlinear physics in the SOL of stellarators. As a result, the Flux-Coordinate-Independent (FCI) method for parallel derivatives~\cite{hariri13a, shanahan16a, hill17a} has been implemented in the BOUT++ framework~\cite{dudson09a}, and an extension to stellarator geometries has been performed within the BSTING project~\cite{shanahan18b, shanahan24a}. While FCI has proven an effective tool for modeling complex topologies~\cite{stegmeir15a,stegmeir16a,body20a,tork26}, it provides additional challenges when treating boundary conditions. Within BOUT++, the Leg-Value-Fill method has been implemented for FCI boundaries~\cite{hill17a}. Additionally, the FCI method relies on interpolation in the perpendicular plane, which adds an extra computational cost and can introduce artificial diffusion~\cite{stegmeir23} which can be prohibitive at low resolutions. Here, we will present the latest developments within the BSTING project which allow for both transport simulations as well as fluid turbulence simulations in the Wendelstein 7-X stellarator.

The paper covers novel developments to the BOUT++ framework in section~\ref{sec:comp}, particularly a new differential operator (\ref{sec:op}), improved parallelization (\ref{sec:parallel}), and an algorithm to ensure physically-consistent treatment of boundary conditions (\ref{sec:bc}), which are traditionally a significant hindrance in FCI methods~\cite{hill17a, stegmeir16a}.  Advancements to zoidberg, the FCI grid generator for BOUT++, are detailed in Section~\ref{sec:grids}. A summary of the results and the implications of this work are provided in Sections~\ref{sec:outlook} and~\ref{sec:conclusions}, respectively.

\section{Performance improvements}\label{sec:comp}
The calculation of nonlinear dynamics in the Wendelstein 7-X SOL has
mandated extensive numerical improvements in order to enhance
numerical efficiency. In this section, we describe several
enhancements which have substantially ameliorated the stability and
efficiency of the FCI methods within BOUT++.
\subsection{Differential operators}\label{sec:op}
The differential operators in BOUT++ in general are derived using a
coordinate transform to compute a derivative on the computational
mesh.  This approach follows the derivation from
D'haeseleer~\cite{dhaeseleer12a,bout510}.
This approach however breaks down if the coordinate transformation is
not differentiable.  This is the case in the presence of an x-point in field-aligned coordinates or
can be the case if a physical target structure
is used as a bounding layer for the domain. In these cases the
transformation contains a corner. While for the rest of the domain the
error scales as expected with the grid spacing, in the presence of the
corner this is not the case.

\begin{figure}
  \centerline{\includegraphics[width=.3\textwidth]{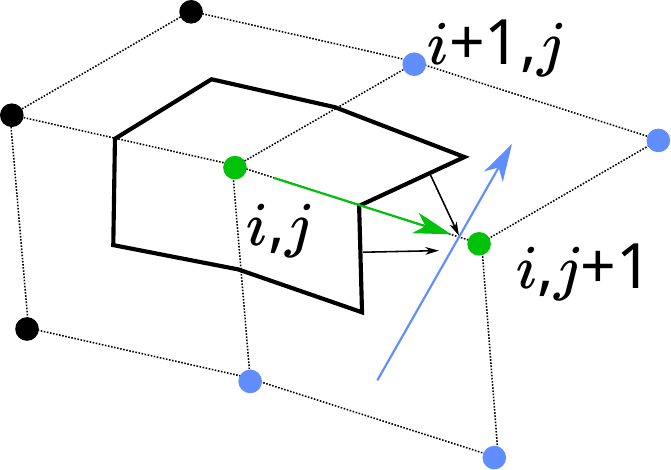}}
  \caption{Example of a cell for the finite volume scheme showing a
    cell. The cell is an octagon around the point of the grid. In
    green are shown the points to construct the gradient across the
    cell boundary, and in blue the points to calculate the other
    gradient contribution. The black arrows denote
    the cell face areas.\label{f:fv_cell}}
\end{figure}

Here an alternative approach for the perpendicular diffusion operator
\begin{align}
  \nabla \cdot a \nabla_\perp f
\end{align}
is proposed, where $a$ is a spatially varying, isotropic diffusion
coefficient and $f$ is a scalar field. This assume that the magnetic field is mostly in toroidal direction, which is in low shear devices like W7-X a good approximation.
In order to conserve the quantity of the field $f$, a finite volume
approach is used. As the magnetic field of stellarators like
Wendelstein 7-X feature low shear, it is assumed that the magnetic field
is purely toroidal, and thus $\nabla_\perp$ lies fully
within an \rzplane.
This reduces the diffusion operation to a set of 2D problems. This is
consistent with previous FCI implementation for BOUT++.
For this finite volume operation the cell considered is shown in
fig.~\ref{f:fv_cell}.
The cell around point $x_{i,j}$ is an octagon consisting of the 8
points $(x_{i,j} + x_{i\pm 1,j}) /2$,  $(x_{i,j} + x_{i,j\pm 1}) /2$
as well as the corners, e.g.
$(x_{i,j} + x_{i + 1,j} + x_{i,j + 1} + x_{i + 1,j +1}) /4$.

In order to calculate the diffusion equation, the gradients across the
8 surfaces have to be reconstructed. For this the normal vectors of the cell faces are
computed, which are shown in fig.~\ref{f:fv_cell} as black arrows.

To obtain the gradient across the surfaces, a linear combination is used.
For example, to get the derivative between $x_{i,j}$ and $x_{i,j+1}$,
one component is in $x_{i,j+1} - x_{i,j}$ direction, shown in green.
The other one,
$(x_{i+1,j} - x_{i-1,j} + x_{i+1,j+1} - x_{i-1,j+1}) / 4$, is
logically orthogonal and shown in blue. A linear system of equations can be solved,
given the two coefficients for the two parameters.

Most of the calculation can be done during grid generation. At simulation
time, only 5 parameters for each cell are needed, namely the volume of the
cell and the radial (green arrow) and perpendicular (blue arrow) flux coefficients in the $x$ and $z$
direction.

The generation is part of zoidberg, the grid FCI generator for BOUT++, and was merged in pull request (PR)
33\footnote{\url{https://github.com/boutproject/zoidberg/pull/33}} and is part
of the latest zoidberg
release\footnote{\url{https://github.com/boutproject/zoidberg/tree/v0.3.0/zoidberg/stencil_dagp_fv.py}}.
The differential operators are part of the Hermes-3 FCI
branch\footnote{\url{https://github.com/boutproject/hermes-3/tree/fci/include/div_ops.hxx}
and \url{https://github.com/boutproject/hermes-3/tree/fci/src/div_ops.cxx}}

\subsubsection{Convergence testing}
\begin{figure*}[t]
  \centerline{\includegraphics[width=\textwidth]{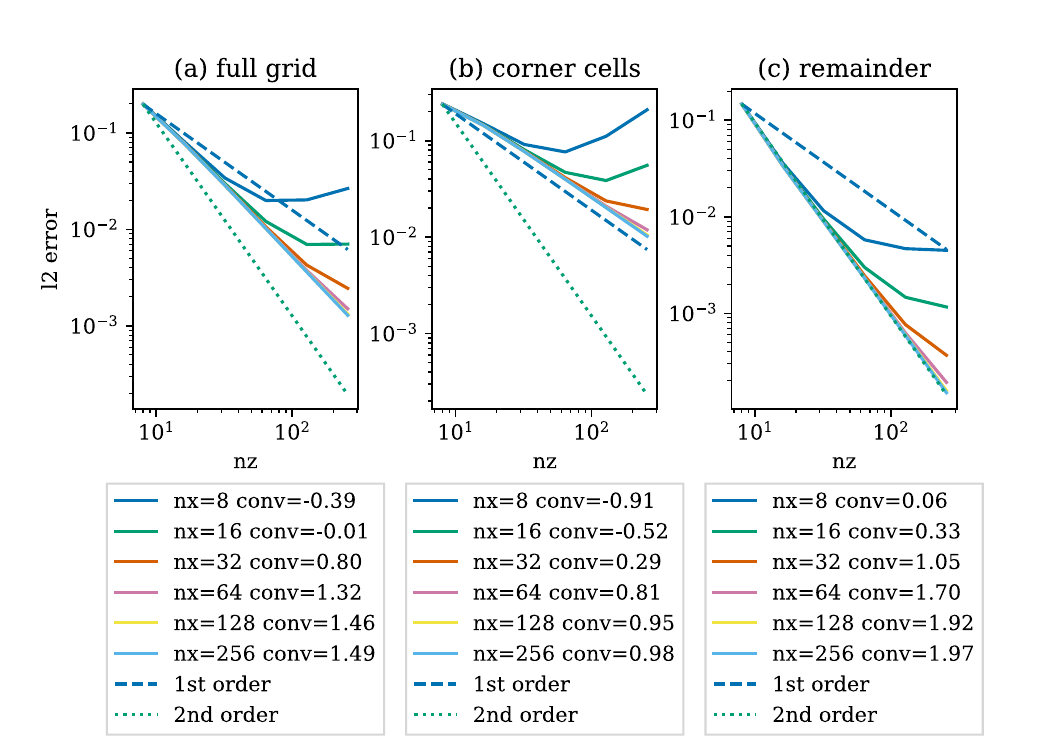}}
  \caption{Convergence plot of a MMS test of the differential
    operator for different radial resolution
    $nx$ and $nz$. The convergence for the small grid scaling is given in as ``conv'' in the legend.\label{f:mms}}
\end{figure*}
\begin{figure*}[t]
  \centerline{\includegraphics[width=.7\textwidth]{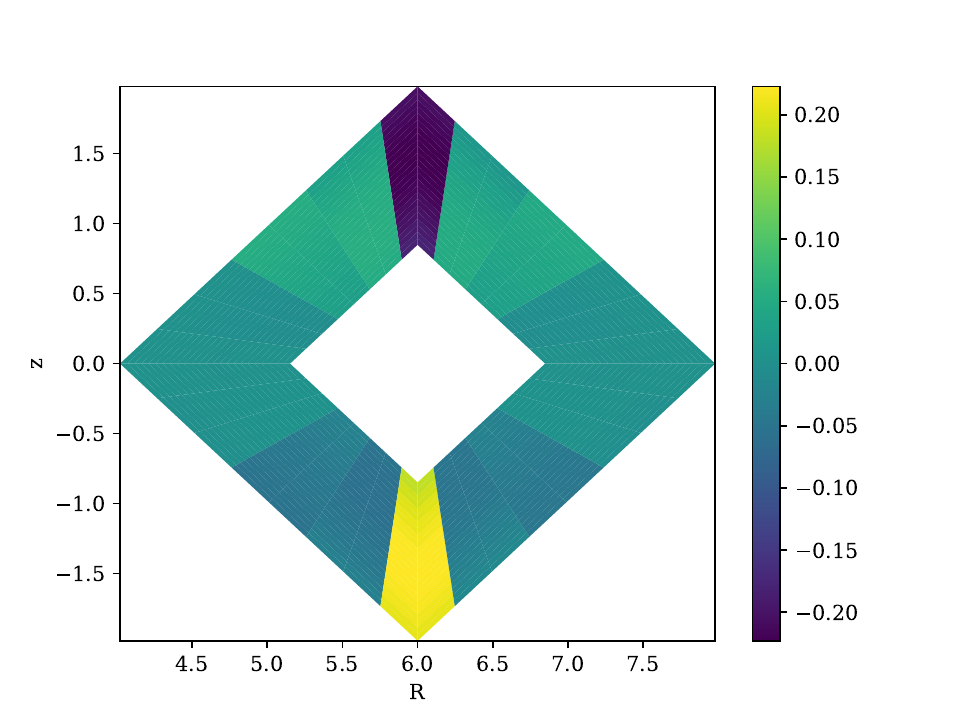}}
  \caption{Error distribution for the MMS plot from fig.~\ref{f:mms} for $nz=nz = 16$. \label{f:mms3}}
\end{figure*}
\begin{figure*}[t]
  \centerline{\includegraphics[width=.48\textwidth]{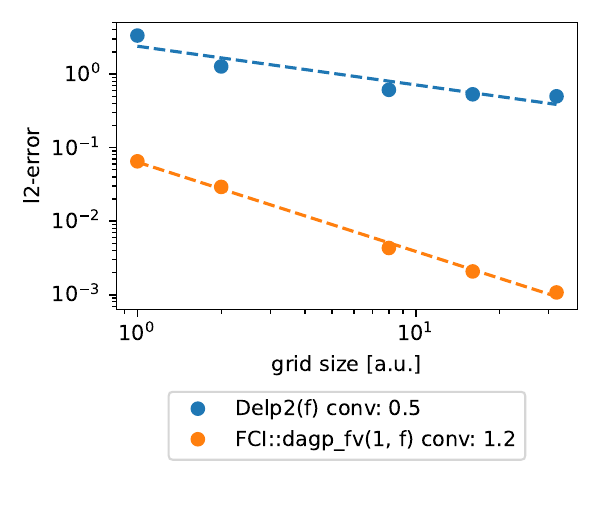}}
  \centerline{\includegraphics[width=.48\textwidth]{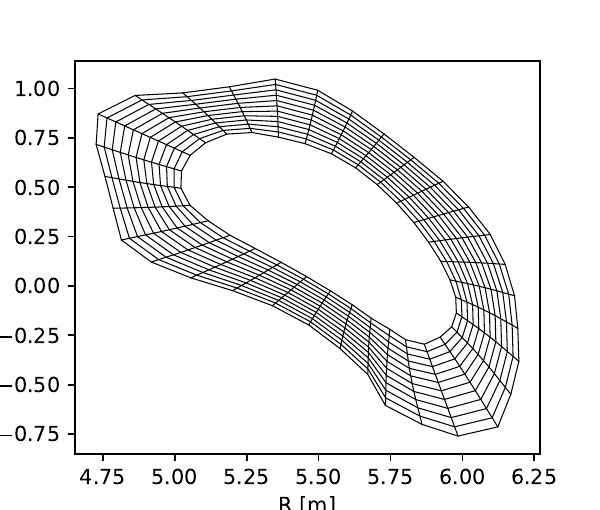}}
  \caption{Convergence plot of a MMS test of the differential operator
    for a W7-X grid and comparison to previous differential operators
    and an example cross section. The grid is refined uniformly in $x$
    and $z$ direction.\label{f:mms2}}
\end{figure*}
The implementation of this differential operator has been tested using the
Method of Manufactured Solutions (MMS)~\cite{roache98a,salari00a}. Thereby the
numerical solution $s$ for a given input function, $\sin(Z)$ is compared to the
analytical solution $a=-\sin(Z)$. This is done for different grid resolutions, and the
error is plotted as a function of the grid spacing.

The l2-error $\sqrt{\sum_{i=1}^N(s_i-a_i)^2 / N}$ is shown in
fig.~\ref{f:mms} and fig.~\ref{f:mms3}.
The computational grid is a hollow square, and thus features corners.
Figure~\ref{f:mms}(a) depicts the l2-error for the full domain, while figure~\ref{f:mms}(b) includes only
the error for the part of the domain that includes the non-smooth part,
i.e. the corners. Figure~\ref{f:mms}(c) shows the l2-error for the smooth part of the domain. It can be seen that for the smooth part  second order conversion is
achieved, which is expected for second-order derivative operators.

For the the corners of the domain, the error reduces only linearly with the grid
spacing, i.e. first order convergence is achieved. As the part of the domain
that includes the corner scales as $\frac{1}{\text{nz}}$ and the total error
is dominated by the corners, the total convergence is in total of order 1.5.

It can also be seen that part of the lines do not show the discussed
convergence. In these cases the error is dominated by the resolution in radial
direction, and reduces with nx. The chosen test function primarily determines at which resolution the error begins to diverge from the expected convergence.

Fig.~\ref{f:mms2} shows the l2-error for a W7-X grid. In addition to the newly
implemented operator, it also shows the convergence of the Delp2 operator, which is an equivalent finite difference operator which has previously been implemented in BOUT++.
In difference to the above mms test, the convergence is reduced, as finer grids contain more structure, and thus new corners are added, reducing the convergance.
It shows that the new implementation has a lower error norm and better convergence. While Delp2 is a finite difference operator, the new operator is a finite volume operator, which is beneficial for numerical stability.

\subsection{Parallelization}\label{sec:parallel}
In the FCI scheme, the calculation of the parallel slices requires interpolation using the values at the neighboring grid points where the field line intersects the next slice. As there is no constraint on where in the computational grid the field line ends, this makes it difficult to implement an efficient communication routine, particularly if the data within a slice are distributed over several MPI tasks.
The original FCI implementation in BOUT++ therefore required that an entire parallel slice remain on one MPI task. This limits MPI parallelization to the different slices in the y-direction and requires that the x-z planes are not split.

An initial attempt at parallelization was carried out using OpenMP. With OpenMP the domain is not decomposed, unlike with MPI. However, the idiomatic BOUT++ physics model needs to be adapted to use outer loops. To achieve a performative OpenMP implementation, the small loops must be merged.
The blob2d model, part of the BOUT++ examples, is a 2 field model that can be used to simulate seeded filaments. It has been rewritten using this technique. Such a change improves performance significantly — for the blob2d model, speedups of around a factor of two have been reported — but it also reduces some of the flexibility of BOUT++, as runtime options must be moved to compile time to avoid branching within a hot loop.
It also requires substantial effort to convert all loops to outer loops, and the numerical derivative operators must be adapted as well.


Instead, the BOUT++ FCI interpolation wes re-implemented as a single linear operator.
In order to do the interpolation, the position where a field line intersects a parallel slice is needed in index space.
The coefficients $t_x$ and $t_z \in [0, 1)$ are the normalized position within the cell in the parallel slice in $x$ and $z$ direction, while $j_x$ and $j_z$ are the integer part of the position.
The original implementation consisted of taking derivatives in grid space of the field $f$:
\begin{align}
  f_x &= \text{DDX}(f)\\
  f_z &= \text{DDZ}(f)\\
  f_{xz} &= \text{DDX}(\text{DDZ}(f))\\
  \intertext{and the hermitian coefficients}
  h_x &= \begin{bmatrix}
    2 t_x^3 - 3 t_x^2 + 1\\
    -2 t_x^3 + 3 t_x^2\\
    t_x (1 - t_x) (1-t_x)\\
    t_x^3 - t_x^2
  \end{bmatrix}
  \intertext{and $h_z$ the same but with $t_z$ instead of $t_x$}
  h&=h_z \otimes h_x
  \intertext{and applying a stencil}
  f_\text{interp}[i] &= h \cdot \cdot
  \begin{bmatrix}
    f[j] & f[j^x] & f_x[j] & f_x[j^x]\\
    f[j^z] & f[j^{xz}] & f_x[j^z] & f_x[j^{xz}]\\
    f_z[j] & f_z[j^x] & f_{xz}[j] & f_{xz}[j^x]\\
    f_z[j^z] & f_z[j^{xz}] & f_{xz}[j^z] & f_{xz}[j^{xz}]
  \end{bmatrix}
\end{align}
with $i$ the index where the data is to be computed, $j=(j_x, j_z)$ the index in the
lower corner where the field line is traced to, $j^x=(j_x+1, j_z)$ is $j$ shifted in $x$ by
+1, similarly $j^z=(j_x, j_z+1)$ and $j^{xz}=(j_x+1, j_z+1)$.

This can however be replaced by a $4\times 4$ stencil, as the derivatives can also
be implemented as stencils.  As a stencil operation is a sparse matrix operation,
the total operation is a sum of matrix-matrix-vector multiplications. As this
is linear, they summation and the matrix-matrix multiplication can be computed
before hand, and at simulation time only the matrix-vector multiplication
remains.

The non-vanishing elements $m$ in a matrix row of $M$ look like
\begin{align}
  m&=\begin{bmatrix}
   h_{2,2} / 4
&
 h_{3,2} / 4- h_{0,2} / 2
&
- h_{1,2} / 2- h_{2,2} / 4
&
- h_{3,2} / 4
\\
 h_{2,3} / 4- h_{2,0} / 2
&
 h_{0,0}+ h_{3,3} / 4- h_{0,3} / 2- h_{3,0} / 2
&
 h_{1,0}+ h_{2,0} / 2- h_{1,3} / 2- h_{2,3} / 4
&
 h_{3,0} / 2- h_{3,3} / 4
\\
- h_{2,1} / 2- h_{2,2} / 4
&
 h_{0,1}+ h_{0,2} / 2- h_{3,1} / 2- h_{3,2} / 4
&
 h_{1,1}+ h_{1,2} / 2+ h_{2,1} / 2+ h_{2,2} / 4
&
 h_{3,1} / 2+ h_{3,2} / 4
\\
- h_{2,3} / 4
&
 h_{0,3} / 2- h_{3,3} / 4
&
 h_{1,3} / 2+ h_{2,3} / 4
&
 h_{3,3} / 4
 \end{bmatrix}
\end{align}
Where $m_{k_x,k_z}$ i multiplied with the field $f_{j_x+k_x-1, j_z+k_z-1}$ in
the parallel slice. The full matrix for applying the interpolation is
$M_{i,j}$ with the $i$-th row containing $m_k$ for the relevant entries and
$0$ otherwise.

This method is implemented in PR 2651 for
BOUT++~\footnote{https://github.com/boutproject/BOUT-dev/pull/2651} and is available in the latest version of BOUT++.  Using
PETSc~\cite{petsc-efficient,petsc-user-ref} this allows to split the domain in
$x$-direction. It also allows splitting in $z$-direction, however that is not
currently supported by BOUT++.
\subsubsection{Scaling}
\begin{figure*}[t]
  \centerline{\includegraphics[width=.5\textwidth]{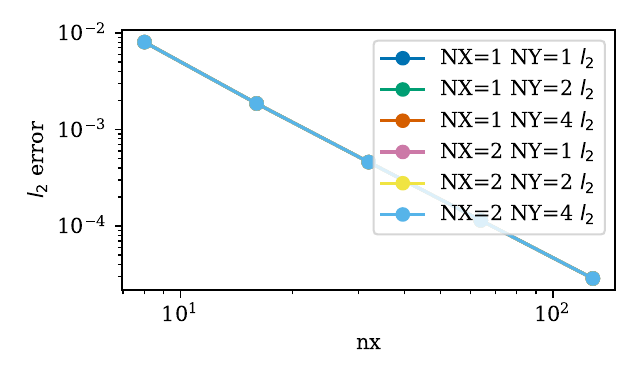}}
  \caption{Scaling of the MMS test of the PETSc parallelized implementation of
    the FCI operator.\label{f:mms:petsc}}
\end{figure*}
The processor-splitting of the FCI domain in the $x$ direction has been
tested. Fig.~\ref{f:mms:petsc} shows the $l_2$ error norm of the FCI operator
as a function of the number of grid points in the $x$ direction for different
numbers of processors in $x$ and $y$ direction. It can be seen that the error
is independent of the number of processors.

\begin{figure*}[t]
  \centerline{\includegraphics[width=.5\textwidth]{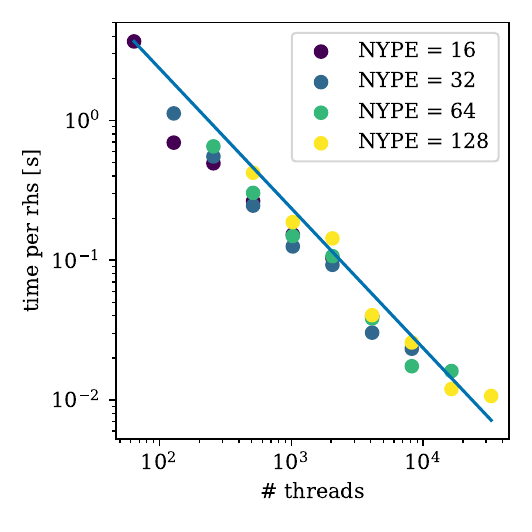}}
  \caption{Scaling of wall clock computation time per evaluation of the right
    hand side of the
    Hermes-2 
    physics
    model. This includes several FCI parallel derivatives and other MPI
    communications, but does not include the time solver and disk i/o. NYPE is the number of MPI decomposition in $y$ direction. Without PETSc, at most 128 threads could be used for the shown grid.}
  \label{f:scaling:petsc}
\end{figure*}
The PETSc parallelized FCI operator enables simulations to run the
$516\times 128\times 512$ grid on up to
32768 MPI processors. The wall time of evaluation of the physics model for
different number of processors is shown in Fig.~\ref{f:scaling:petsc}.
For 32768 MPI processors the domain on each processor is
$2\times1\times 512$ without guard cells or $6\times 3 \times 512$
with guard cells. As BOUT++ currently does not support domain
decomposition in $z$-direction, for this processor count most of the
fields are guard cells, limiting the scaling.

\subsection{Boundary conditions for FCI}\label{sec:bc}
The boundary condition implemented uses the Leg-Value-Fill (LVF)
method~\cite{hill17a}, which uses a Taylor expansion to extrapolate to the
boundary, including the value on the boundary, as well as values in the
domain. The LVF method uses a Taylor expansion around the point in the boundary of a given
order. Constraining the Taylor expansion for the values in the domain, as well
as either the value or the derivative at the point of the derivative, 
gives a linear system of equations.
In the current implementation, this linear system of equations is solved
analytically by sympy~\cite{sympy} using code generation before compilation. This provides a direct solution in a computationally
efficient manner at run time, while at the same time reducing the likelihood
of introducing bugs during calculation or implementation of the expressions.

\begin{figure*}[t]
  \centerline{\includegraphics[width=.5\textwidth]{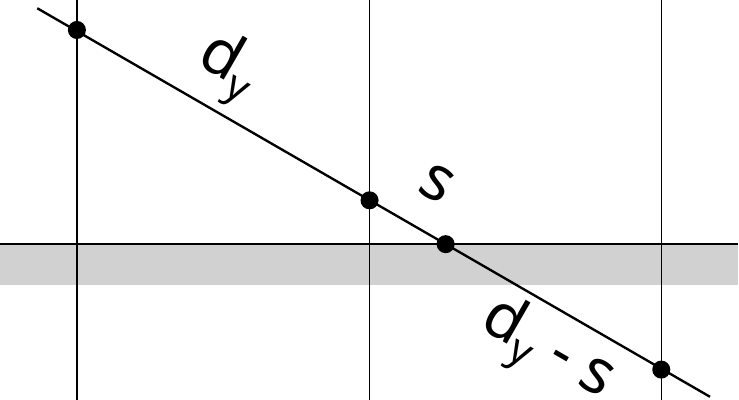}}
  \caption{Plot of a field line leaving the domain. The grid spacing $d_y$ and the distance $s$ between the last point in the domain and the boundary is shown.}
  \label{f:bc:dys}
\end{figure*}
Care has to be taken if the distance $s$ between last point in the domain and the
boundary is small in the case of Dirichlet boundary conditions. Fig.~\ref{f:bc:dys} shows the lengths.
Care has to be taken when the
distance $s$ is much smaller then the parallel grid spacing $d_y$, then any
difference in value between the boundary $f_b$ and the last point in the domain $f_d$
makes the value in the boundary diverge as $f_d + d_y \frac{f_b-f_d}{s}$ for a second order boundary condition.
Thus a cut-off for
$s/d_y=0.01$ is used, after which the value in the domain is ignored to avoid
divergence, and the code falls back to a lower order scheme.

Another issue in the case of fully 3D boundaries is the presence of short
parallel connection lengths. They are in general always present unless
special care is taken to avoid them. While the grid generation reduces the
amount of short connection length regions, they are not fully avoided, as
discussed in section~\ref{ss:zb:short}. Thus the computational framework should
be able to handle them. In the case of higher order boundary conditions,
values from the previous slide are used to extrapolate into the boundary.
In the case of a short connection length, a point in the domain has
boundaries on both sides, thus higher order schemes cannot be
used. Rather then requiring that all cases use a lower order scheme,
the user can request a high order scheme, and the code falls
automatically back to an appropriate, lower order scheme only where it
is required.

This is implemented as part of PR 3308\footnote{\url{https://github.com/boutproject/BOUT-dev/pull/3308}}.

\subsection{Custom boundary conditions}\label{sec:ubc}
While BOUT++ provides a method to set boundary conditions simple cases
using the input file, more complicated boundary conditions, like
sheath boundary conditions, have to be implemented in the physics
module.

In the case of a field aligned grid, this can be
implemented using boundary iterator provided by BOUT++. The here shown
example uses an index (\code{i}) for the last point in the domain, one
(\code{im}) for the first point in the boundary and one (\code{ip})
for the previous point in the bulk.

Similar code can be written for FCI, that has to use different
boundary iterators, as the sheath boundary is no longer a logically
rectangular domain. Care has also be taken to fall back to lower order
schemes, if the previous point is also a boundary, as discussed in
sec.~\ref{sec:bc}.

This becomes significantly more complicated if higher-order schemes are used in the parallel direction. When using only one parallel slice, the boundary is always between the evolved slice and the parallel slice.
With more than one parallel slice, the boundary may be between the evolved slice and the first parallel slice, or it may be between the parallel slices. Depending on that, boundary conditions may need to be applied to all parallel slices or just some. While this is certainly possible, the resulting code is lengthy and hard to read. For example, switching to a third order extrapolation for the Dirichlet boundary condition in the last line becomes somewhat tedious, as it needs to write out the stencil and additionally needs branching, depending on how many parallel slices can be used. A new approach has therefore been implemented. This provides a common
interface for both boundary regions. To provide a unified API,  many of the details requiring special attention for FCI are abstracted behind the interface.

The sheath boundary physics code for the FA version is shown in
lst.~\ref{lst:fa}. The FCI version is shown in lst.~\ref{lst:fci}.
The new version, which is capable of handling both FCI and FA, is shown in
lst.~\ref{lst:uni}. 
The unified code is clearer and more concise because it hides all the details of the parallel gradient scheme.

This allows to use the boundary operators discussed in
sec.~\ref{sec:bc} for FCI, including checks whether the field line leaves
the domain immediately. It also includes the correct handling of
the leg-fill method, which was missing from the implementation in lst.~\ref{lst:fci}. It uses
a template function, that is always instantiated for both FA and FCI.
This allows the compiler to optimize FA and FCI independently. At
runtime, both versions are available, and thus it is a runtime
decision whether FCI or FA is used.

Writing unit tests for the implementation of the member functions such as
\code{extrapolate\_grad\_o2} and \code{dirichlet\_o2} is much easier
than writing tests for the entire physics implementation.  Having a
unified code for FA and FCI means only one test for the
physics is needed. This approach also ensures that fixes for one version benefit both versions.

The BOUT++ part that enables this is implemented in PR 3308 \footnote{\url{https://github.com/boutproject/BOUT-dev/pull/3308}}.
This is used in Hermes-3 in the fci branch \footnote{\url{https://github.com/boutproject/hermes-3/blob/fci/src/sheath_boundary_parallel.cxx}}.

\section{Stellarator elliptic grid generation}\label{sec:grids}
The grid can significantly impact on the simulation. For example, the result may be inaccurate due to large cells or strong
deformation. Depending on the numerical method used, the sensitivity to the various effects can vary. Poor grid quality can sometimes be compensated for by using a finer mesh. However, in this case, a poor grid can increase the computational demand
significantly because a finer grid means more cells to compute. It also means that the grid spacing is smaller and smaller structures can be
resolved. For a given wave speed, that implies that shorter time scales also
need to be resolved, which further increases the computational cost~\cite{courant28a}.

\subsection{Short connection length}\label{ss:zb:short}
\begin{figure*}[t]
  \centerline{\includegraphics[height=9pc]{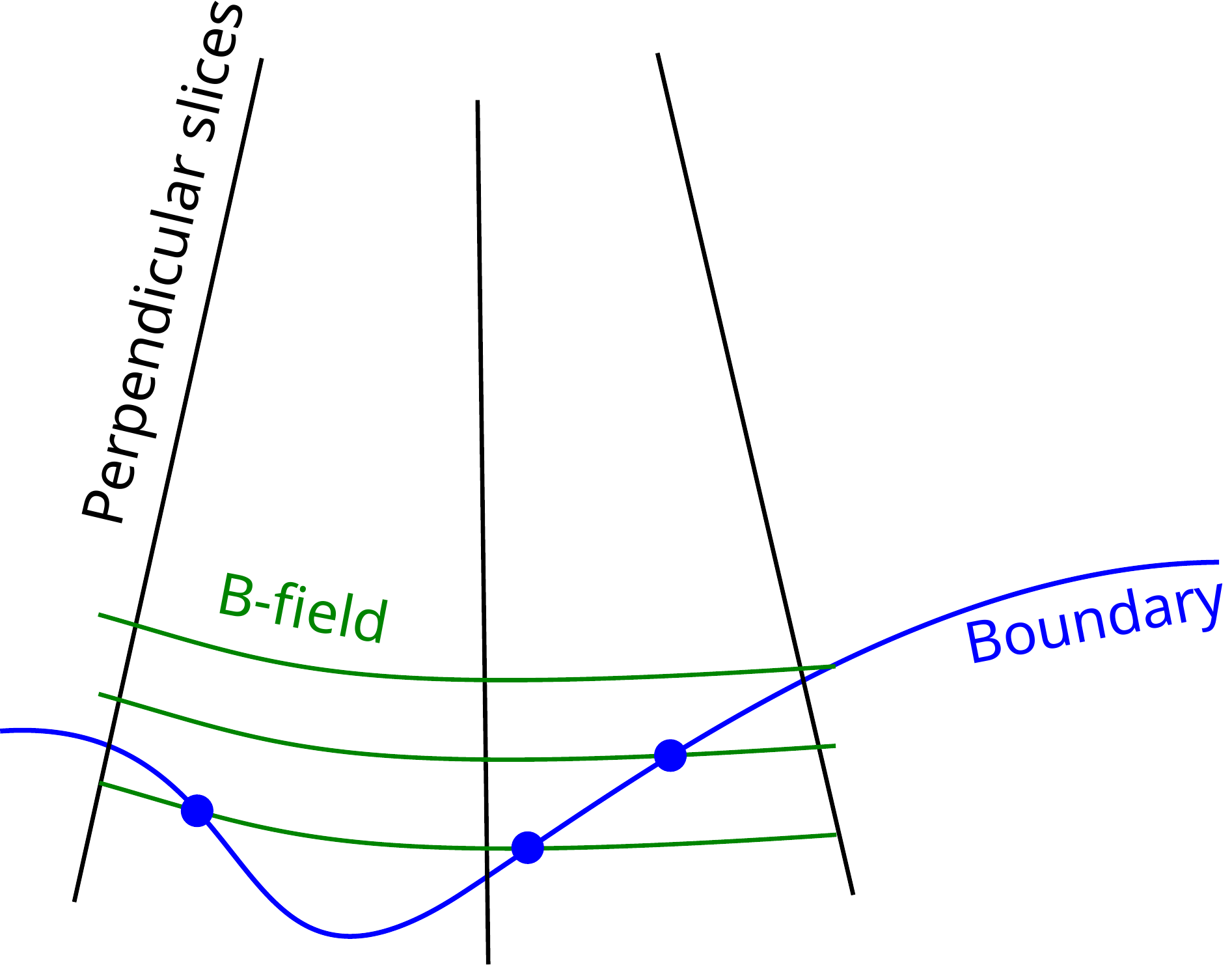}}
  \caption{Sketch of a field line entering the domain just before the slice
    and leaving the domain just after the slice.\label{f:slc}}
\end{figure*}

Using the first wall as the boundary for the computational domain
results in areas where the field lines enter and leave the domain
adjacent to the parallel slice, resulting in boundaries on both
sides. This is sketched in fig.~\ref{f:slc}.  These regions of short
connection length are not expected to contain significant plasma
because such short connection lengths would inherently lead to
excessive loads.

From a computational standpoint, they are not particularly interesting either because the parallel operators have boundary conditions on both sides, so the region is largely defined by the boundary conditions. This can lead to areas
where numerical issues arise due to the strong impact of the boundary conditions.

\begin{figure*}[t]
  \centerline{\includegraphics[height=9pc]{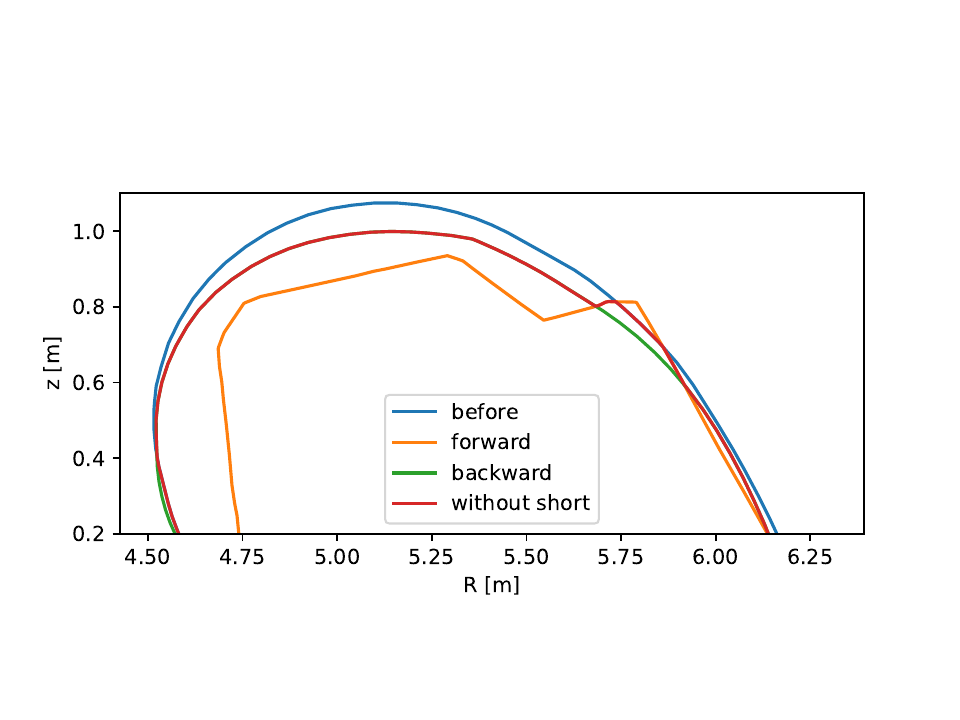}}
  \caption{Plot of a cross section, including the tracing from the previous and next cross section. The area that is not covered by any of the slices is removed.\label{f:scl2}}
\end{figure*}
Therefore, it is important to reduce the part of the computational domain that
consists of areas of short connection length. To achieve this, each outer
boundary is traced one slice in each direction. Thus for each slice the
parallel projections of the previous and next slices are available. The removal is implemented using shapely~\cite{shapely} where the boundaries
are represented by polygons. First, the union of the traced polygons is
computed; then, the intersection of the union and the local polygon is
computed. This in turn gives the updated outer boundary. An example is shown in fig.~\ref{f:scl2}.
Using this method,
some short field lines may remain, which can be attributed to the finite
precision of the tracing, as well as due to the different representation of the
outer boundary as a polygon or a spline.
However, the remaining areas can be addressed with appropriate boundary
conditions, see sec.~\ref{sec:bc}.

This is implemented in zoidberg-w7x and is available in the latest release\footnote{\url{https://github.com/dschwoerer/zoidberg-w7x/blob/v1.2.1/zoidberg_w7x/zoidberg_examples.py\#L512}}.

\subsection{Alignment of boundaries}\label{sec:align}
The poloidal grid generation is easier if the points which lie on the inner and outer boundary are aligned, providing a nearly-orthogonal mesh. In the previous version of zoidberg this was achieved by aligning the
outer most point. This works for simple shapes quite well, but fails however in
the presence of strong shaping or additional structures on the outer
boundary, such as ports.

In the current zoidberg version, two new methods have been
implemented. The first minimizes the total distance between the inner
and outer points.

The second method starts with projecting the points of the outer boundary onto
the inner boundary. The poloidal position of the points on the inner
boundary is smoothed with a Gaussian filter. This ensure that the points are
in monotonic increasing order. A user-provided option allows to set the
maximum ratio of smallest to largest poloidal distance. Allowing a larger
variation gives a more orthogonal mesh.
The grids shown here are
examples for W7X grids in the standard configuration. The resolution is
$(nx, ny, nz) = (68, 36, 256)$.

\begin{figure*}[t]
  \centerline{\includegraphics[width=\textwidth]{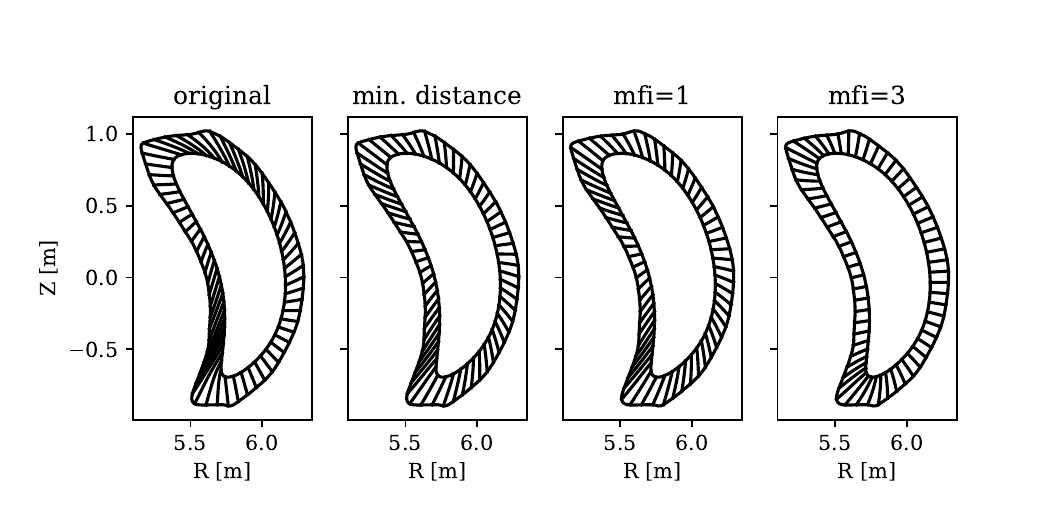}}
  \caption{Example grids, depending on the alignment of the
    inner and outer boundary. Shown is the inner and outermost poloidal line,
    as well as every forth radial grid line. On the left is the ``original'' implementation
    that has been used previously to this work. ``min. distance'' minimizes
    the total distance between the respective points for inner and outer
    points. ``mfi=1'' and ``mfi=3'' are both based on the orthogonal
    projection method, but with different values for the allowed poloidal
    variation of the grid spacing on the inner surface. While ``mfi=1'' allows
    for no variation in the spacing, ``mfi=3'' allows for up to a factor of 3,
    i.e. the largest poloidal distance is 3 times larger then the smallest
    poloidal distance on the inner surface.\label{f:align:spatial}}
\end{figure*}
Fig.~\ref{f:align:spatial} shows the effect of the poloidal alignment on the
grid quality. It can be seen that for the ``original'' implementation most of
the cells are significantly sheared. The effect is smallest at the radially
outermost point, where the lines have been aligned. The ``min. distance''
method and the orthogonal projection method with ``mfi=1'' show virtually the
same behavior. ``mfi=3'' allows for more variation in the poloidal spacing on
the inner boundary. This results in a more orthogonal mesh.

\begin{figure*}[t]
  \centerline{\includegraphics[width=.4\textwidth]{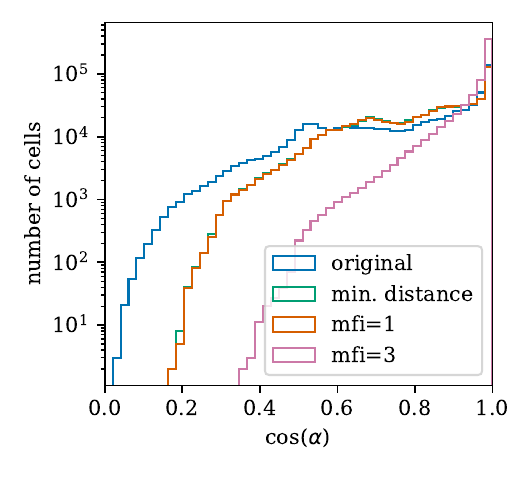}}
  \caption{Histogram of the orthogonality of the grid depending on the
    poloidal alignment method. For each cell of the grid the cosine of the angle between
    the $dx$ and the $dz$ derivative are computed. See
    fig.~\ref{f:align:spatial} for a description of the different methods
    shown.\label{f:align:norm}}
\end{figure*}
Fig.~\ref{f:align:norm} shows the histogram of the orthogonality of the grid
depending on the poloidal alignment method. It can be seen that for the
original method several cells are close to purely parallel. Such cells are
numerically challenging, as calculating derivatives perpendicularly requires
subtracting two roughly parallel vectors, which increases the requirement for
accuracy of those derivatives. For
``min. distance'' and ``mfi=1'', the minimum angle found is significantly
reduced. The two methods are again very similar, confirming what has been
shown in fig.~\ref{f:align:spatial}. For the orthogonal projection method with
an allowed poloidal variation, ``mfi=3'', not only the smallest observed
value of cos$(\alpha)$ is increased, but also the distribution is much
stronger peaked around 1 and drops of more quickly.

This is implemented in zoidberg in PR 23 \footnote{\url{https://github.com/boutproject/zoidberg/pull/23}}.

\subsection{Distribution within domain}\label{sec:distribution}
In previous works the elliptic grid generation was used~\cite{shanahan18b,shanahan24a}.
This results in smooth grids, as perturbations on the boundaries are dampened.
However, it pushes the cells towards the inner boundary. This leads to
larger cells near the outer boundary and smaller cells near the inner
boundary. Especially near outwards pointing features on both the inner and outer boundary, that effect is particularly strong.

The equation for the grid solved by zoidberg has been changed to
\begin{align}
  0 &=
  \alpha \frac{\partial^2 \vec r}{\partial^2x^2}
  - 2 \beta \frac{\partial^2 \vec r}{\partial^2 x z}
  + \gamma \frac{\partial^2 \vec r}{\partial^2 z^2}
  \intertext{with}
  \alpha &=
  \zeta^2 \left(\frac{\partial R}{\partial x}\right)^2 +
  \left(\frac{\partial Z}{\partial x}\right)^2\\
  \beta &=
  \zeta^2 \frac{\partial R}{\partial x}
  \frac{\partial R}{\partial z} +
  \frac{\partial Z}{\partial x}
  \frac{\partial Z}{\partial z}\\
  \gamma &=
  \zeta^2 \left(\frac{\partial R}{\partial z}\right)^2 +
  \left(\frac{\partial Z}{\partial z}\right)^2
\end{align}
where $\zeta$ is a coefficient that relaxes the derivatives in $z$ direction. In
the current version $\zeta = 10/3$ is used. This seems to give a good
compromise between smooth contours away from the boundaries, as well as
uniformity in cell size. For $\zeta = 1$ the previous implementation
is restored.

\begin{figure*}[t]
  \centerline{\includegraphics[width=\textwidth]{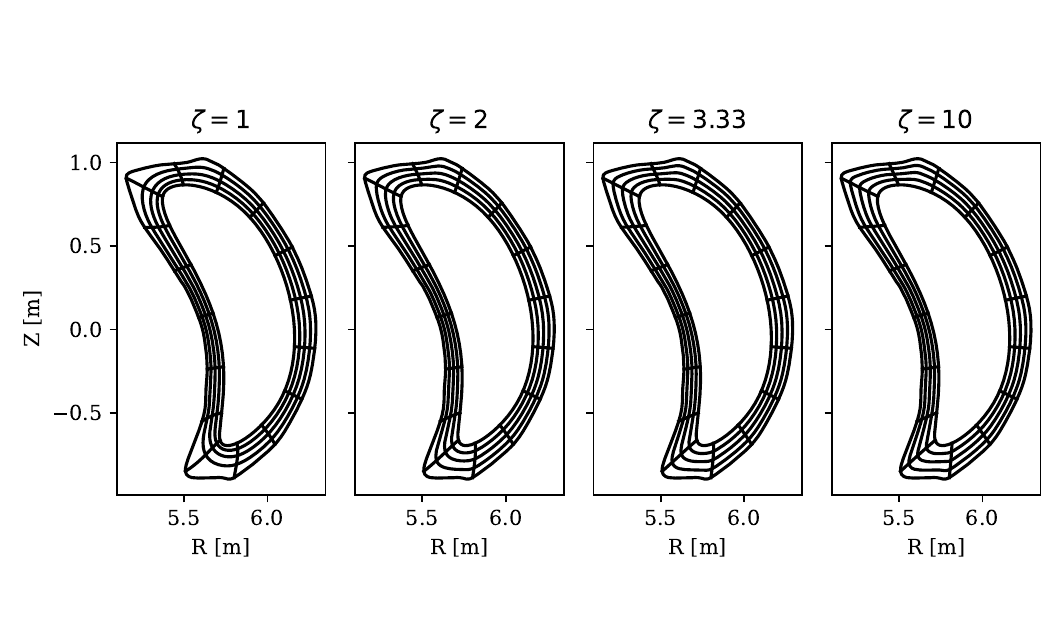}}
  \caption{Plot of the distribution within the domain for different values of
    $\zeta$. With increasing $\zeta$ the cells in the corner get
    smaller. Shown is every 16th line for a $64\times 256$ W7-X grid. The difference is largest at the outward point features on the left top and left bottom. For $\zeta=1$ the outer cells are much larger then the inner cells, while for $\zeta=10$ they are similar sized. The $\zeta=1$ to $\zeta=2$ shows the largest difference.\label{f:relax:spatial}}
\end{figure*}
\begin{figure*}[t]
  \centerline{\includegraphics[width=.4\textwidth]{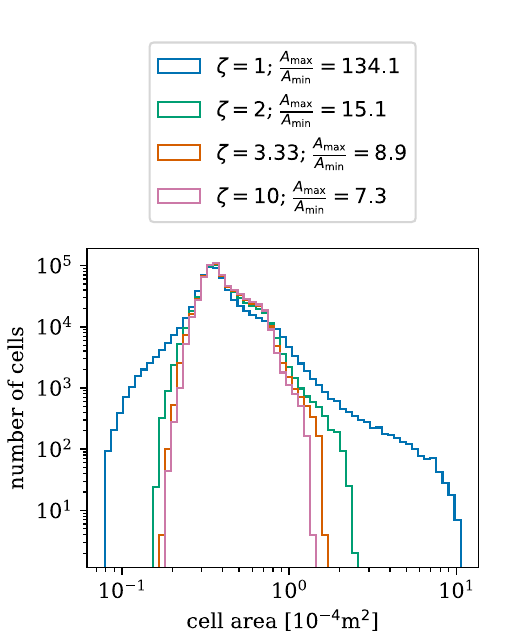}}
  \caption{Histogram of the cell area for different values of $\zeta$. With
    increasing $\zeta$ the cells get more uniform and the ratio of largest to
    smallest cell reduces.\label{f:relax:hist}}
\end{figure*}
Fig.~\ref{f:relax:spatial} shows the spatial distribution of the cells for
different values of $\zeta$ for the modified elliptic grid generation code.
It can be seen that the cell contours get closer to the cell corners for
increasing $\zeta$, as well as that they are less squeezed around the corners
on the inner boundary.
This can also be confirmed in the histogram in fig.~\ref{f:relax:hist}.

This is implemented in zoidberg in PR 33 \footnote{\url{https://github.com/boutproject/zoidberg/pull/33}}.

\subsection{Data source}\label{sec:source}
The code zoidberg-w7x\footnote{\url{https://github.com/dschwoerer/zoidberg-w7x}}.~\cite{zoidbergw7x114}, which includes extensions specific for grids for
Wendelstein 7-X (W7X), has been extended to load data from the
webservices~\cite{bozhenkov13a,grahl20a}.
This allows to use a realistic first wall geometry. This also allows to
easily get updated geometries if, for example, the divertor is changed.

In addition to the first wall, also the field line tracer~\cite{bozhenkov13a} and VMEC
equilibria~\cite{grahl18a} can be used via the webservices.

In addition also a routine has been implemented that uses an EMC3-Eirene~\cite{feng14a,feng21a} grid
as input. This is implemented using the xemc3 code~\cite{xemc3-0.1.0,bold22a} and is intended to
simplify comparison to EMC3-Eirene, as a similar outer boundary can be used in
the simulations.

\subsection{Smoothing of the boundary}\label{sec:smooth}
Some methods have been implemented to smooth the outer boundary.
While those methods could also be applied for the inner boundary, that
is typically not
needed as for the inner boundary often a flux surface from vmec is used.
The outer boundary however can be significantly more complicated.

The boundaries are represented internally by RZline, which uses the scipy
function \code{scipy.interpolate.splrep}~%
\cite{dierckx75a,dierckx82a,dierckx81a,dierckx93a}.
It turned out to be useful to allow
restricting the order to $k=1$, i.e. piecewise linear functions that are not
continuously differentiable to avoid loops and overshoots around sharp
edges. In addition the improved RZline in zoidberg allows to reduce the number
of knots used. This gives a less accurate spline, but smoother spline
representation of the data.

\begin{figure*}[t]
  \centerline{\includegraphics[width=.4\textwidth]{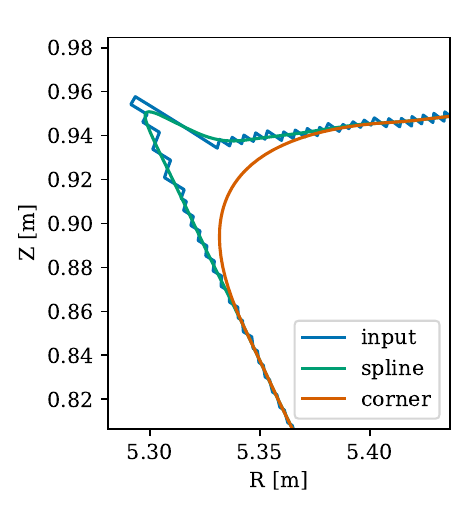}}
  \caption{Example of the applied smoothing method. The input shows the non-smooth input. In this case it is an EMC3-Eirene grid. The spline smoothing can remove the small-scale zig-zag. The "corner" algorithm detects sharp corners, and replaces them by a smoother cubic function, leaving the rest of the boundary unaffected.\label{f:smooth}}
\end{figure*}

One of the issues of the outer boundary is that the domain can include some
area of the pumping gap. This leads to very peaked outer boundaries. They are
causing significant trouble for the grid quality. A routine has been
implemented that detects corners, and removes problematic points in the
proximity of the corner and replaces them by a smooth interpolation, which is shown in fig.~\ref{f:smooth}. This uses a cubic polynomial,
that is constrained to have the same derivative on the edges of the function
as the input data. In addition the cubic contribution is minimized, as the
polynomial is otherwise under-defined.

A different method for smoothing is to use spectral functions. The points $r_i$
on the RZline are approximated by a spectral function of order
$N$ with
\begin{align}
  r_i &\approx \sum_{j=0}^N s_j \sin(j*\theta_i) + c_j \cos(j*\theta_i)
\end{align}
where $\theta_i$ is the toroidal coordinate of $r_i$ and $s_j$ and $c_j$ are
the coefficients, determined by fitting.
Due to the nature of the trigonometric functions, this ensure that the
resulting curve is smooth. Lower values of $N$ give smoother results, which
are in turn computationally more efficient, while higher values of $N$ give
a higher-fidelity boundary.
\begin{figure*}[t]
  \centerline{\includegraphics[width=\textwidth]{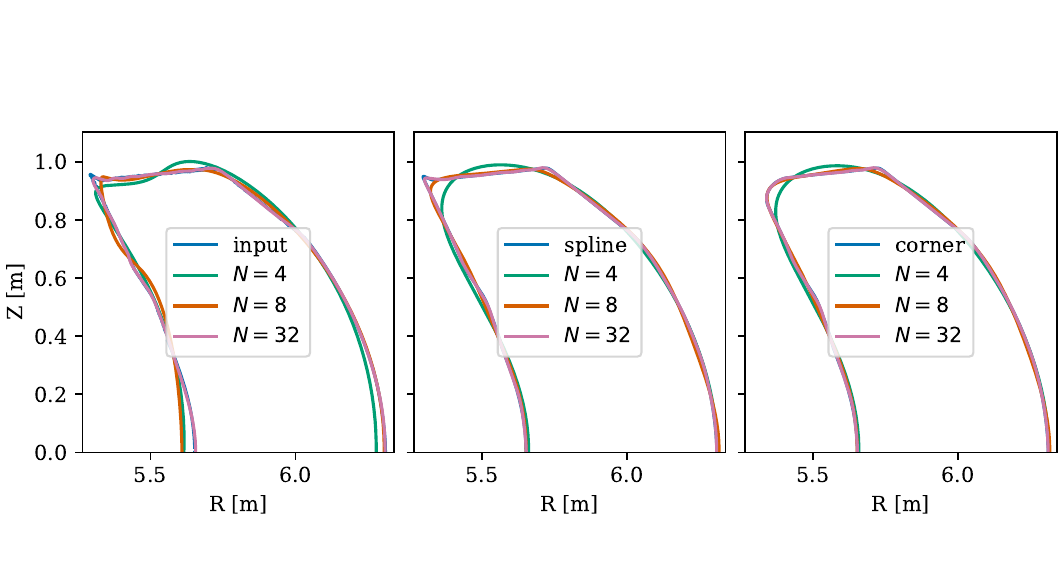}}
  \caption{Spectral smoothing applied to the different boundaries from fig.~\ref{f:smooth}. 
  It can be seen that for a smoother input, the spectral smoothing gives better results.\label{f:smooth:fft}}
\end{figure*}
\begin{figure*}[t]
  \centerline{\includegraphics[width=\textwidth]{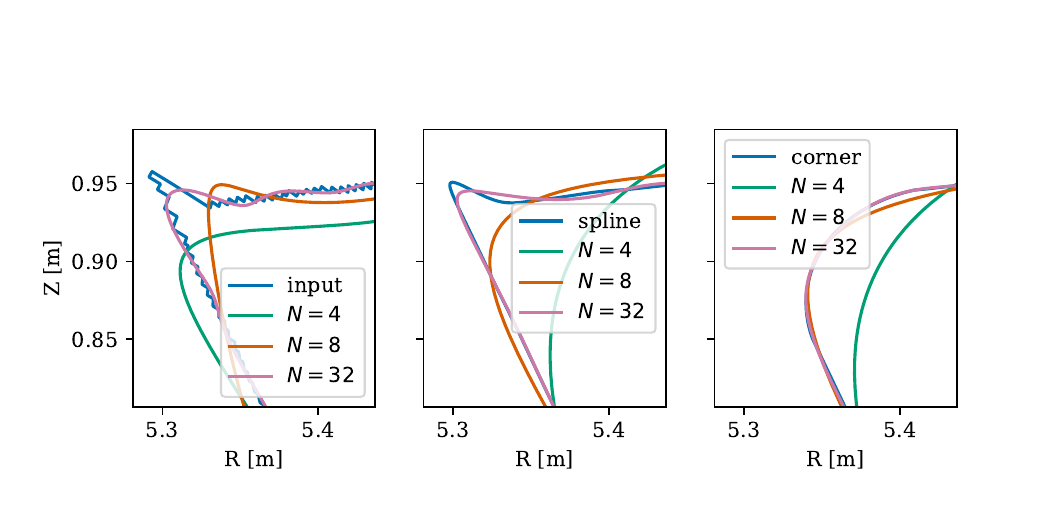}}
  \caption{Zoom in for fig.~\ref{f:smooth:fft}. Only for the "corner" case, the high order ($n=32$) spectral smoothing changes little and does not introduce oscillations.}
\end{figure*}

This is implemented in zoidberg as part of PR 20 \footnote{\url{https://github.com/boutproject/zoidberg/pull/20}}.

\section{Outlook}\label{sec:outlook}
The changes shown here are being used to run Hermes-2 simulations using the
BOUT++ framework with EMC3-like physics and geometry. This will be used as a
test to ensure the correct implementation of the physics. Once this is
done, further terms will be switched on and the impact will be studied. One
example is the impact of drifts, that has been suggested to be of significant
importance in W7-X~\cite{bold24a,flom23a}.

Further work will include adapting the Hermes-3 code~\cite{dudson24a} to FCI,
which allows to benefit from the enhancements with respect to Hermes-2 and continuous improvements from the actively developed code.

Besides the comparison to EMC3, both unit tests for several FCI specific features, as well as MMS tests of operators and the physics in Hermes-3 are currently implemented. This will increase the confidence in the code and avoid regressions in the future, as the tests will be part of the BOUT++ and Hermes-3 continuous integration (CI).

\section{Conclusions}\label{sec:conclusions}
Some significant steps in enabling BOUT++ to compute SOL simulations
of W7-X have been achieved.

The grid generation, using zoidberg, have been improved. New methods
for smoothing the domain boundary have been implemented. Also the
placements of the cells within the domain has been improved by new
methods for align outer and inner boundary, as well as by using a
modified elliptic grid generation scheme.

A new differential operator for $\nabla \cdot a \nabla_\perp f$ was
implemented, that is based on a finite volume approach. Using
pre-computation of the coefficients, that allows for a fast and
accurate computation of this term.

The parallelization of the FCI implementation in BOUT++ has been
improved, allowing for a good scaling to a large number of cores.
Also the boundary condition have been adopted to deal with the
features found in stellarator grids. In addition to the standard
boundaries provided by BOUT++, also a new interface, to implement
custom boundaries like sheath physics, has
been added. This allows to have one implementation for both FCI
and FA in a concise way.

\begin{figure*}[t]
  \centerline{\includegraphics[width=.5\textwidth]{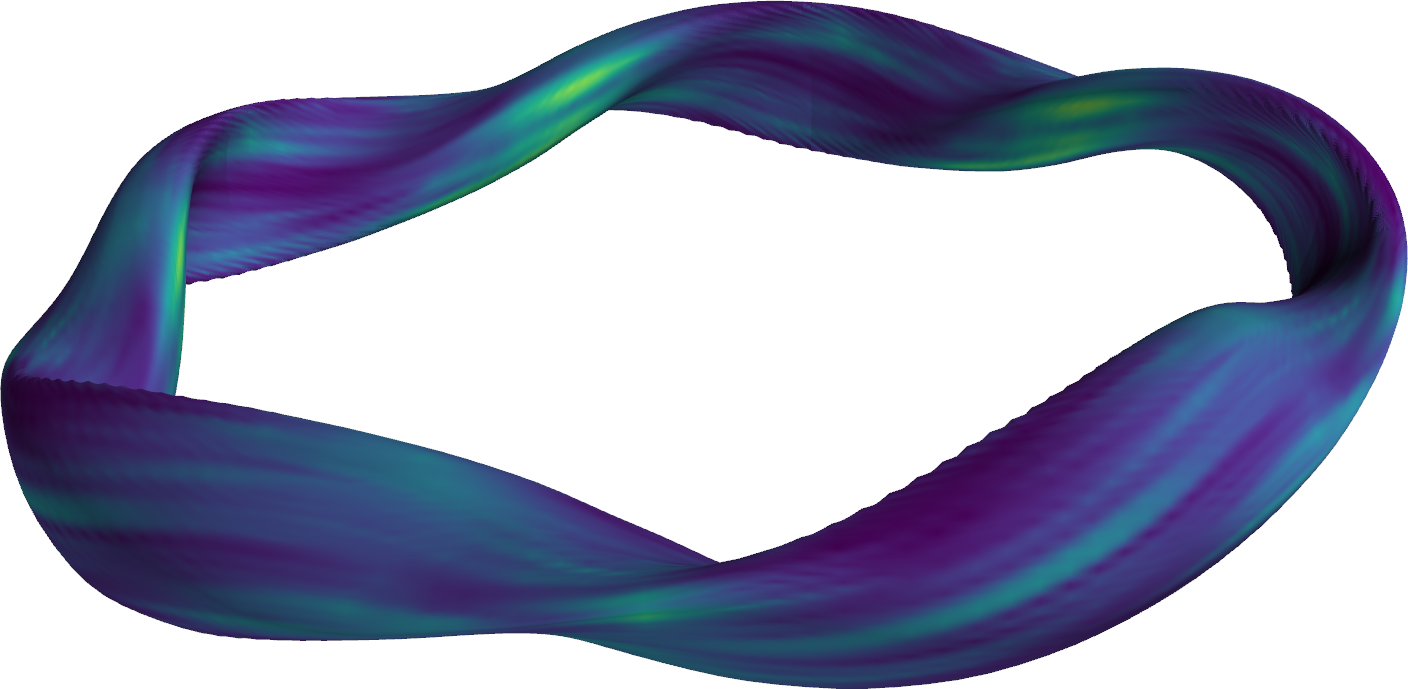}}
  \caption{Density fluctuation level of a turbulence simulation with reduced
    magnetic field.\label{f:turb}}
\end{figure*}
Fig.~\ref{f:turb} shows a plot of the standard deviation of the density of a
turbulence simulation in W7-X geometry using a reduced magnetic field. The
simulations have been running stable and have reached the saturated phase of
the turbulence. This has been published in ref.~\cite{shanahan24a}.

\begin{figure*}[t]
  \centerline{\includegraphics[width=.5\textwidth]{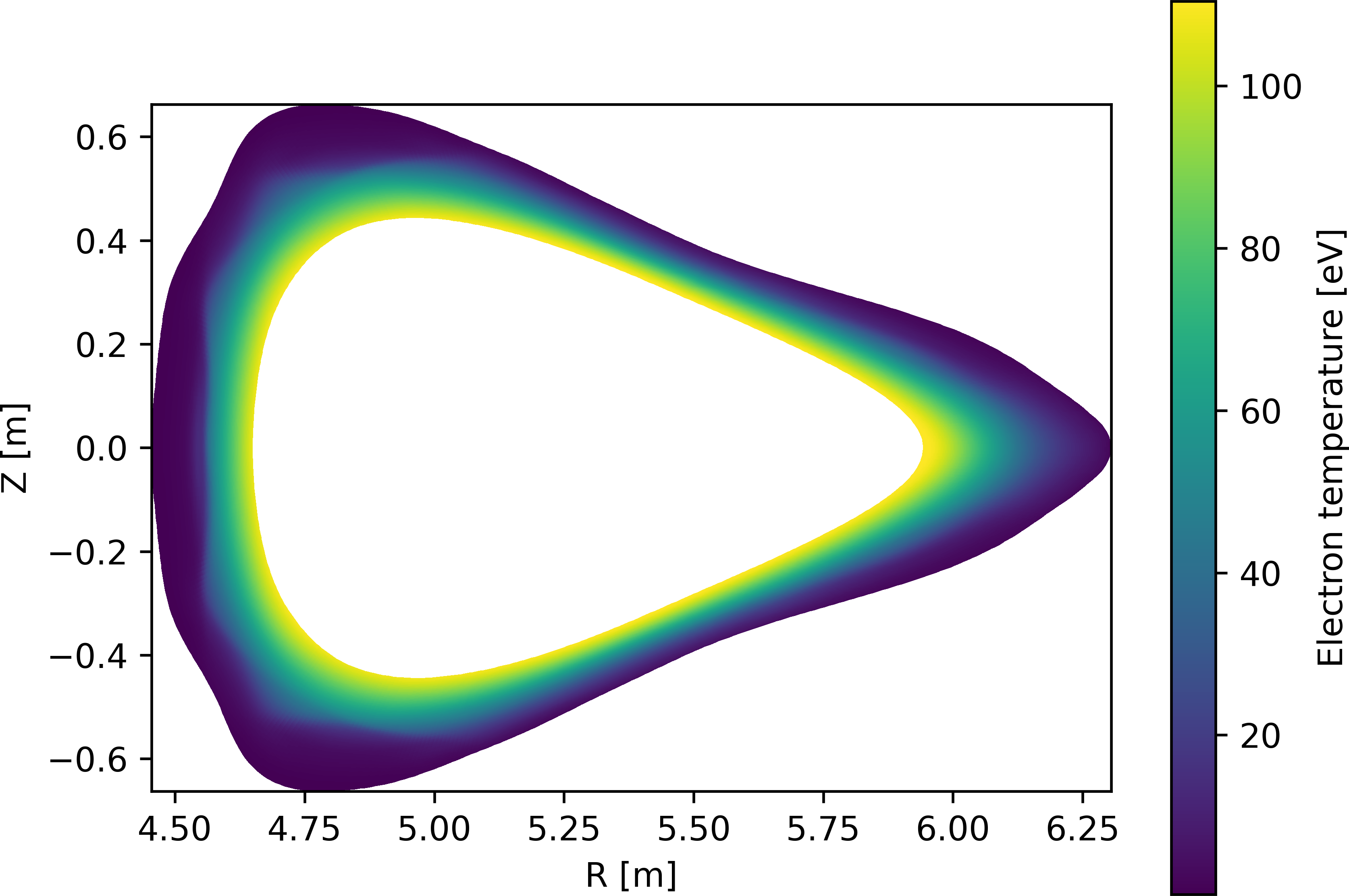}}
  \caption{Plot of the electron temperature in an EMC3 like transport
    simulation.\label{f:emc3}}
\end{figure*}
Fig.~\ref{f:emc3} shows the electron temperature profile of a simulation in
steady state using the Hermes-2 model.
A more detailed comparison with matching EMC3 simulations is outside the scope of this work, the physics of which will be published in a separate manuscript.

\FloatBarrier

\section{Acknowledgement}
This work has been carried out using the xarray
framework~\cite{hoyer17a,xarray_2025_01_1}.
Both xbout~\cite{xbout_038} and xemc3~\cite{xemc3_1_0_0,bold22a} have been used.
Some task have been parallelized using GNU parallel~\cite{tange23a}.

This work has been supported by the German Federal Ministry of Education and Research as part of the funding programme ‘Fusion 2040 - Research on the way to the fusion power plant’ under the funding code 13F1001A.

This work has been carried out within the framework of the EUROfusion Consortium, funded by the European Union via the Euratom Research and Training Programme (Grant Agreement No 101052200--EUROfusion). Views and opinions expressed are however those of the author(s) only and do not necessarily reflect those of the European Union or the European Commission. Neither the European Union nor the European Commission can be held responsible for them.

\bibliographystyle{elsarticle-num}
\bibliography{phd}

\appendix
\FloatBarrier

\section{Custom Boundary Condition Code}
\begin{listing}
\begin{minted}{c++}
if (upper_y) {
  for (RangeIterator r = mesh->iterateBndryUpperY(); !r.isDone(); r++) {
    for (int jz = 0; jz < mesh->LocalNz; jz++) {
      auto i = indexAt(Ni, r.ind, mesh->yend, jz);
      auto ip = i.yp();
      auto im = i.ym();
      double tesheath = Te[i];

      double grad_ne = Ne[ip] - Ne[i];
      [...]
      Ve[im] = 2 * vesheath - Ve[i];
    }
  }
}
\end{minted}
\caption{Sketch of a sheath boundary code for field aligned grids.}
\label{lst:fa}
\end{listing}

\begin{listing}
\begin{minted}{c++}
if (outer_x) {
  for (auto& bndry : mesh->getBoundariesPar(BoundaryParType::xout)) {
    for (auto& pnt : bndry) {
      double ti = Ti[pnt.ind()];
      double grad_ne = 0;
      if (pnt.valid() > 0) {
        grad_ne = Ne.ynext(pnt.dir)[pnt.ind().yp(pnt.dir)] - Ne[pnt.ind()];
      }
      [...]
      Ve.ynext(-pnt.dir)[pnt.ind().ym(pnt.dir)] = 2 * vesheath - Ve[pnt.ind()];
    }
  }
}
\end{minted}
\caption{Sketch of a sheath boundary code for FCI grids, same physics as lst.~\ref{lst:fa}}
\label{lst:fci}
\end{listing}

\begin{listing}
\begin{minted}{c++}
iter_boundary([&](auto& pnt) {
  double tesheath = pnt.interpolate_sheath_o1(Te);
  double grad_ne = pnt.extrapolate_grad_o2(Ne);
  [...]
  pnt.dirichlet_o2(Ve, vesheath);
}
\end{minted}
\caption{Sketch of a sheath boundary code for both FCI and FA, same physics as lst.~\ref{lst:fa} and lst.~\ref{lst:fci}.}
\label{lst:uni}
\end{listing}
\end{document}